\documentclass[sigconf,natbib=true]{acmart}
\usepackage{bigstrut}
\usepackage{booktabs}
\usepackage{multirow}
\usepackage{microtype}
\usepackage{graphicx}
\usepackage{subfig}
\usepackage{overpic}
\usepackage{balance}
\usepackage{amsthm}
\usepackage{array}
\usepackage{clrscode}
\usepackage{multicol}
\usepackage{float}
\usepackage{newfloat}
\usepackage{color}
\usepackage{xcolor}
\usepackage{amsopn}
\usepackage{mathrsfs}
\usepackage{mathtools}
\usepackage{amsmath}
\usepackage{arydshln}
\usepackage{blkarray}
\usepackage{enumerate}
\usepackage{courier}
\usepackage{rotating}
\usepackage{bm}
\usepackage{wrapfig}
\usepackage{ragged2e}
\usepackage[misc]{ifsym}
\usepackage{threeparttable}
\usepackage{makecell}
\usepackage{adjustbox} 
\usepackage{enumitem}
\usepackage{hyperref}
\usepackage{url}
\usepackage{xspace}
\usepackage{comment}
\usepackage{changepage}
\usepackage{algorithm}
\usepackage{algpseudocode} 
\usepackage{algorithmicx} 
\usepackage[ruled, vlined, nofillcomment, linesnumbered, algo2e]{algorithm2e}

\definecolor{codegreen}{rgb}{0,0.3,0.6}
\definecolor{codegray}{rgb}{0.5,0.5,0.5}

\newcommand{\ie}{\emph{i.e.,}\xspace}
\newcommand{\eg}{\emph{e.g.,}\xspace}

\newcommand{\paratitle}[1]{\vspace{1.5ex}\noindent\textbf{#1}}
\newcommand{\wrt}{w.r.t.\xspace}

\newcommand{\ignore}[1]{}

\AtBeginDocument{%
  \providecommand\BibTeX{{%
    \normalfont B\kern-0.5em{\scshape i\kern-0.25em b}\kern-0.8em\TeX}}}

\setcopyright{acmlicensed}
\copyrightyear{2018}
\acmYear{2018}
\acmDOI{XXXXXXX.XXXXXXX}

\acmConference[SIGIR '25]{Proceedings of the 48th International ACM SIGIR Conference on Research and Development in Information Retrieval}{July 13--18, 2025}{Padua, Italy}
\acmISBN{978-1-4503-XXXX-X/18/06}

\begin{document}

\title[Pre-training Generative Recommender with Multi-Identifier Item Tokenization]{Pre-training Generative Recommender with \\ Multi-Identifier Item Tokenization}

\settopmatter{authorsperrow=4}
\author{Bowen Zheng\textsuperscript{*}}
\orcid{0009-0002-3010-7899}
\affiliation{%
    \institution{
    Gaoling School of Artificial Intelligence,  
    Renmin University of China}
    \city{Beijing}
    \country{China}
}
\email{bwzheng0324@ruc.edu.cn}

\author{Enze Liu\textsuperscript{*}}
\orcid{0009-0007-8344-4780}
\affiliation{%
    \institution{
    Gaoling School of Artificial Intelligence,
    Renmin University of China}
    \city{Beijing}
    \country{China}
}
\email{enzeliu@ruc.edu.cn}

\author{Zhongfu Chen}
\affiliation{%
    \institution{Poisson Lab,\\ Huawei}
    \city{Beijing}
    \country{China}
}
\email{chenzhongfu3@huawei.com}

\author{Zhongrui Ma}
\affiliation{%
    \institution{Poisson Lab,\\ Huawei}
    \city{Beijing}
    \country{China}
}
\email{zhongrui.ma@huawei.com}

\author{Yue Wang}
\affiliation{%
    \institution{Poisson Lab,\\ Huawei}
    \city{Beijing}
    \country{China}
}
\email{wangyue262@huawei.com}

\author{Wayne Xin Zhao
\textsuperscript{\Letter}
}
\orcid{0000-0002-8333-6196}
\affiliation{
    \institution{
    Gaoling School of Artificial Intelligence,  
    Renmin University of China}
    \city{Beijing}
    \country{China}
}
\email{batmanfly@gmail.com}

\author{Ji-Rong Wen}
\orcid{0000-0002-9777-9676}
\affiliation{
    \institution{
    Gaoling School of Artificial Intelligence,  
    Renmin University of China}
    \city{Beijing}
    \country{China}
}
\email{jrwen@ruc.edu.cn}

\thanks{* \ Equal contribution.} 
\thanks{\Letter \ Corresponding author.} 

\renewcommand{\shortauthors}{Bowen Zheng, et al.}

\begin{abstract}

Generative recommendation has emerged as a promising paradigm that recommends the potential item by autoregressively generating its identifier.
Most existing methods adopt a strict \emph{one-to-one mapping} strategy, where each item is represented by a single identifier.
However, this rigid tokenization scheme poses issues, such as suboptimal semantic modeling for low-frequency items and limited diversity in token sequence data.

To overcome these limitations, we propose \textbf{MTGRec}, which leverages \underline{M}ulti-identifier item \underline{T}okenization to augment token sequence data for \underline{G}enerative \underline{Rec}ommender pre-training.
Our approach is built upon two core innovations: \emph{multi-identifier item tokenization} and \emph{curriculum recommender pre-training}.
For multi-identifier item tokenization, we leverage the Residual-Quantized Variational AutoEncoder (RQ-VAE) as the tokenizer backbone and treat model checkpoints from adjacent training epochs as semantically relevant tokenizers. 
This allows each item to be associated with multiple identifiers, enabling a single user interaction sequence to be converted into several token sequences as different data groups.
For curriculum recommender pre-training, we introduce a curriculum learning scheme guided by data influence estimation. 
Specifically, we estimate the influence of each tokenizer’s data using first-order gradient approximation and dynamically adjust the sampling probability of each data group during recommender pre-training. 
After pre-training, we fine-tune the model using a single tokenizer to ensure accurate item identification for recommendation. Extensive experiments on three public benchmark datasets demonstrate that MTGRec significantly outperforms both traditional and generative recommendation baselines in terms of effectiveness and scalability.
Our code is available at \textcolor{blue}{\url{https://github.com/RUCAIBox/MTGRec}}.

\end{abstract}

\ccsdesc[500]{Information systems~Recommender systems}

\keywords{Generative Recommendation, Item Tokenization}

\maketitle

\section{Introduction}
\label{sec:introduction}

Nowadays, sequential recommender systems~\cite{gru4rec,fpmc} have been widely used in various online platforms, aiming to capture users' personalized preferences based on their historical interaction behaviors.
Traditional sequential recommendation approaches~\cite{bert4rec,gru4rec,sasrec,caser} assign a unique ID to each item and predict the next item by measuring the similarity between the user preference and candidate items through approximate nearest neighbor (ANN) algorithms.
Most recently, driven by the promising potential of large language models (LLMs)~\cite{llm_survey} and generative retrieval methods~\cite{dsi,nsi,genret,autoindexer}, several studies proposed applying the generative paradigm as an alternative to ANN in recommender systems~\cite{gptrec,tiger,howtoindex,lc-rec,mbgen}. 
The core idea of generative recommendation lies in employing a list of tokens (\ie \emph{a token sequence}) as the identifier for item representation instead of a single vanilla ID. 
Thus, the next-item prediction is reformulated as a sequence-to-sequence problem, aiming to autoregressively generate the identifier of the target item.

A typical generative recommendation framework consists of two key components, namely the \emph{item tokenizer} and the \emph{generative recommender}. 
The item tokenizer is crafted to associate each item with a list of tokens that implies semantic knowledge. 
The merit is that shared tokens between items reflect the underlying semantic similarities.
Existing methods are developed by utilizing a variety of techniques such as co-occurrence matrix decomposition~\cite{gptrec,howtoindex}, hierarchical clustering~\cite{seater,eager}, and multi-level codebook~\cite{tiger,letter,tokenrec}. 
Among these, the Residual-Quantized Variational AutoEncoder (RQ-VAE)\cite{rqvae} is the most commonly used item tokenizer. 
And recent researches attempt to further improve the quality of item identifiers by incorporating collaborative signals~\cite{letter,etegrec} or multi-behavior information~\cite{mbgen}. 
The generative recommender is leveraged to autoregressively generate the target token sequence, usually employing either decoder-only (\eg GPT~\cite{gpt2,gpt3}) or encoder-decoder (\eg T5~\cite{t5}) architectures. 
Furthermore, some studies focus on enhancing the generative recommender by using dual decoders~\cite{eager} or incorporating contrastive learning~\cite{seater}.

Despite remarkable progress, previous methods typically represent each item by a single identifier, employing a strict \emph{one-to-one mapping} for item tokenization, which leads to the following two potential issues.
First, token sequence data inherits the long-tail distribution and data sparsity issues of interaction data~\cite{sgl,s3rec}. 
As a result, tokens associated with long-tail items are low-frequency and lack supervision signals, making it challenging to effectively learn their semantics.
Second, the one-to-one mapping restricts the diversity of sequence data.
Compared to all possible token permutations, mapping observed item sequences to token sequences in a one-to-one manner results in a lack of variation.
Moreover, these limitations hinder the potential for performance improvement through model scaling, as observed in LLMs~\cite{llm_survey,scaling_law,scaling_instruct}.

In light of these issues, our idea is to associate one item with multiple identifiers, which is achieved by incorporating multiple item tokenizers with semantic relevance.
The advantages of this multi-identifier scheme are two-fold.
First, associating each item with more tokens increases the exposure frequency of tokens and promotes token sharing across items, thereby facilitating the effective learning of token semantics.
Second, an item interaction sequence can be tokenized into several token sequences, thereby enriching the diversity of the training data.
In addition, the increased volume and diversity of token sequence data enable us to achieve performance improvement through model scaling.
To develop our approach, we focus on two key challenges: (i) learning multiple semantically relevant rather than extraneous item tokenizers; and (ii) effectively training a generative recommender based on the proposed multi-identifier item tokenization.

To this end, we propose a novel framework namely \textbf{MTGRec}, which incorporates \underline{M}ultiple item \underline{T}okenizers to improve the effectiveness and scalability of the \underline{G}enerative \underline{Rec}ommender.
Overall, our approach associates each item with multiple identifiers and augments one item sequence into several token sequences, which serve as training data for generative recommender pre-training.
Specifically, we focus on two key aspects, namely \emph{multi-identifier item tokenization} and \emph{curriculum recommender pre-training}.
For multi-identifier item tokenization, we adopt the learnable RQ-VAE as the backbone and take the model checkpoints corresponding to adjacent epochs as semantically relevant item tokenizers.
In this way, we can tokenize each item to multiple identifiers and construct several groups of token sequence data.
Each group is characterized by related but distinct distributions derived from different item tokenizers.
For curriculum recommender pre-training, we design a data curriculum scheme to adaptively schedule the proportions of these data groups during model pre-training.
As the specific technique, we measure the influence of data from each item tokenizer using first-order gradient approximation and adjust the sampling probability of each data group accordingly.
Finally, we fine-tune the pre-trained model using a single item identifier to ensure precise item identification during recommendation.

In summary, our primary contributions are as follows:

$\bullet$ We propose a novel framework \textbf{MTGRec} that learns multiple item tokenizers for curriculum recommender pre-training to improve generative recommendation.

$\bullet$ We develop a multi-identifier item tokenization approach for token sequence augmentation and introduce a data curriculum scheme based on data influence estimation to enhance recommender training.

$\bullet$ We conduct extensive experiments on three public datasets, demonstrating the superiority of our proposed framework over both traditional and generative recommendation baselines in terms of effectiveness and scalability.

\section{Methodology}
\label{sec:methodology}

In this section, we present our proposed generative recommender \textbf{MTGRec}, which uses multi-identifier item tokenization to augment token sequences for generative recommender pre-training.

\begin{figure*}[]
\centering
\includegraphics[width=0.98\linewidth]{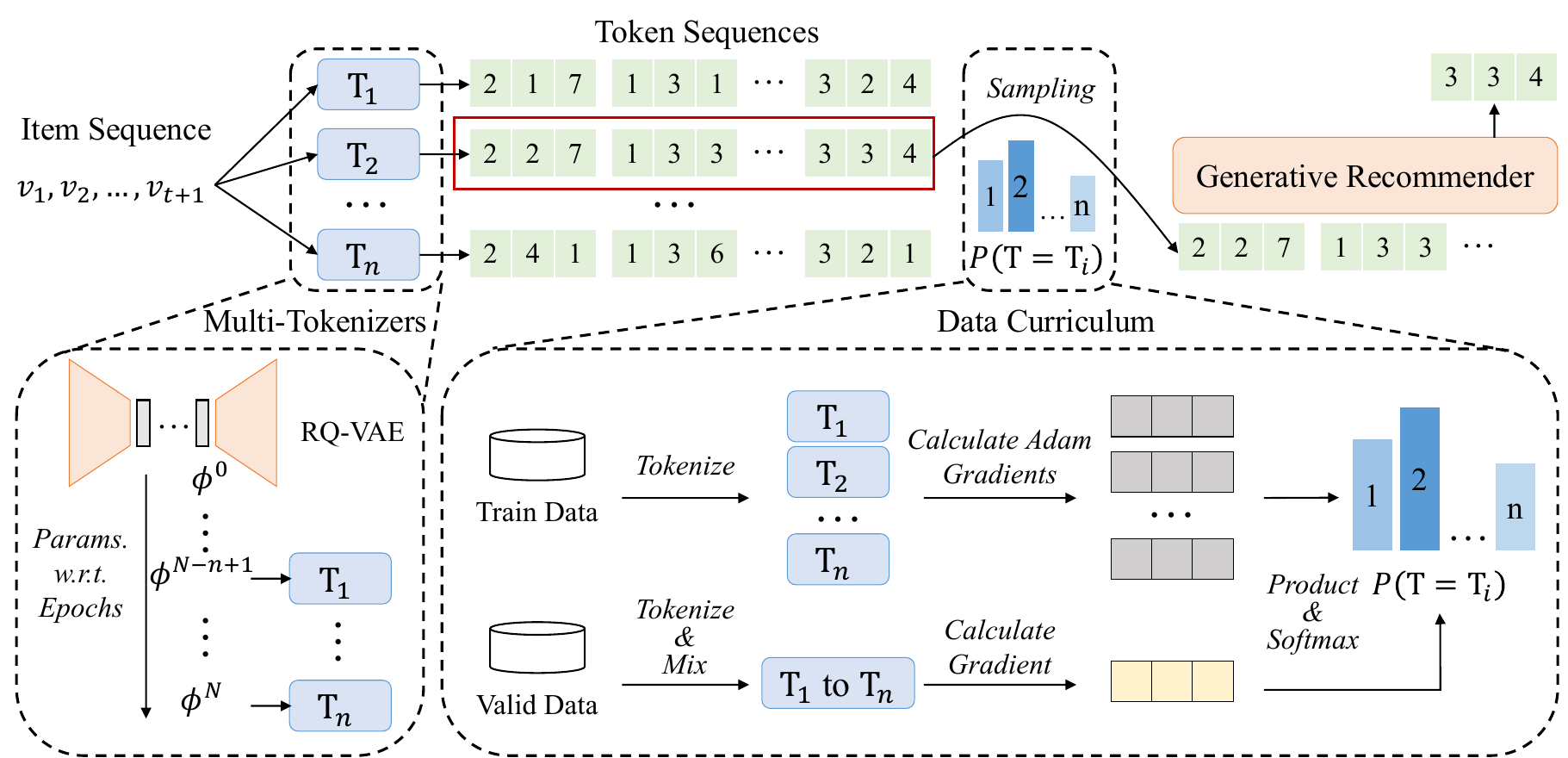}
\caption{The overall framework of MTGRec with two key techniques. (i) We utilize RQ-VAE checkpoints of adjacent epochs as semantically relevant item tokenizers and tokenize an item sequence into multiple token sequences. (ii) We propose a data curriculum scheme based on data influence estimation, which is implemented through first-order gradient approximation.}
\label{fig:model}
\end{figure*}

\subsection{Overview}
\label{sec:overview}

\subsubsection{Problem Formulation}
Given the item set $\mathcal{V}$, let $S = [v_1, \dots, v_t]$ denote the user's historical interacted items in chronological order.
Sequential recommendation aims to capture user preferences implicit in the item sequence and predict the next potential item $v_{t+1}$.
Generative recommendation reformulates the traditional sequential recommendation task as a sequence-to-sequence problem. 
In this paradigm, an item tokenizer $\mathrm{T}$ is learned to represent each item by a token sequence as its identifier.
Formally, we refer to the above process as item tokenization, denoted as $[c_1,\dots,c_H] = \mathrm{T}(v)$, where $c_h$ represents the $h$-th token of $v$ and $H$ is the length of the identifier.
Then, the interacted item sequence $S$ and the target item $v_{t+1}$ are tokenized into token sequences $X = \mathrm{T}(S) = [c^1_1,c^1_2, \dots, c^t_{H-1},c^t_H]$  and $Y = \mathrm{T}(v_{t+1}) = [c^{t+1}_1,\dots,c^{t+1}_H]$ respectively, where each item is represented by $H$ tokens.
Finally, the next-item prediction is achieved by autoregressively generating the identifier of the target item (\ie $Y$).
Formally, this task can be written as:
\begin{align}
    P(Y|X) = \prod_{h=1}^H  P(c^{t+1}_h|X,c^{t+1}_1,\dots,c^{t+1}_{h-1}).
\end{align}

\subsubsection{Method Overview}

Different from previous works~\cite{tiger,letter} that establish a \emph{one-to-one mapping} between each item and its identifier, our idea is to associate one item with multiple identifiers to construct more massive and diverse token sequence data for recommender pre-training. 
To this end, we make efforts in the following two aspects:

$\bullet$ \textbf{Multi-Identifier Item Tokenization}~(Section~\ref{sec:seq_aug}):
We employ the learnable RQ-VAE~\cite{rqvae} as the backbone of the item tokenizer. 
To obtain item tokenizers with semantic relevance, we select multiple RQ-VAE checkpoints corresponding to adjacent epochs during the training process.
Applying these tokenizers, each item is associated with multiple identifiers, allowing a single item sequence to be tokenized into several token sequences. 
These token sequences, generated by semantically relevant tokenizers, encapsulate related yet distinct semantic knowledge.

$\bullet$ \textbf{Curriculum Recommender Pre-training}~(Section~\ref{sec:data_curr}): 
Given the hybrid data derived from different item tokenizers, we propose a curriculum learning method to adaptively schedule the proportions of these data groups during model pre-training.
Specifically, we design a data curriculum scheme that increases the proportion of useful data while decreasing the proportion of low-quality data.
To achieve this, we utilize the first-order gradient approximation to estimate the data influence, measuring whether it is ``useful'', and dynamically adjust the sampling probabilities of different data groups accordingly.
Finally, we fine-tune the pre-trained model based on a single item identifier to ensure precise item identification.

The overall framework of the proposed approach is shown in Figure~\ref{fig:model}. Next, we will present the details of our method.

\subsection{Multi-Identifier Item Tokenization}
\label{sec:seq_aug}
As described above, we propose a multi-identifier scheme to tokenize one item sequence to several token sequences: (i) employ the learnable RQ-VAE as the backbone of item tokenizer (Section~\ref{sec:rqvae}), (ii) select semantically relevant tokenizers from adjacent epochs (Section~\ref{sec:rel_tokenizers}), and (iii) tokenized the item sequence by multiple item tokenizers (Section~\ref{sec:tokenize_seq}).

\subsubsection{Tokenizer Backbone}
\label{sec:rqvae}
In practice, we implement the item tokenizer as a RQ-VAE~\cite{rqvae}, which is advantageous due to its efficacy in modeling item semantics and mitigating length bias~\cite{letter,seater}.
Initially, RQ-VAE takes the item semantic embedding $\bm{z}$ (\eg text embedding encoded by a pre-trained language model) as input and encodes it into a latent representation $\bm{r}$.
Then, $\bm{r}$ is quantized into serialized codes~(called tokens) from coarse to fine through $H$-level residual quantization.
Each level's codebook is denoted by $\mathcal{C}^h = \{\bm{e}_k^h\}_{k=1}^K$, where $\bm{e}_k^h$ is a learnable cluster center and $K$ is the codebook size. 
Finally, the residual quantization is applied to $\bm{r}$:
\begin{align}
    & c_h = \underset {k} { \operatorname {arg\,min} } ||\bm{r}_h - \bm{e}_k^h||_2^2, \\
    & \bm{r}_{h+1} = \bm{r}_h - \bm{e}_{c_h}^h,
    \label{eq:rq}
\end{align}
where $\bm{r}_h$ is the residual vector in the $h$-th RQ level, and $\bm{r}_1 = \bm{r}$. 
Thereafter, we obtain the item quantized representation $\tilde{\bm{r}}=\sum_{h=1}^H\bm{e}^h_{c_h}$ and feed it into a decoder to reconstruct the item semantic embedding.
Overall, the loss of the RQ-VAE is $\mathcal{L}_{\mathrm{T}} = \mathcal{L}_{\text{recon}} + \mathcal{L}_{\text{rq}}$, where $\mathcal{L}_{\text{recon}} = ||\bm{z} - \hat{\bm{z}}||_2^2$, and $\mathcal{L}_{\text{rq}} = \sum_{h=1}^{H} ||\operatorname{sg}[\bm{r}_h] - \bm{e}_{c_h}^h||_2^2 + \beta \ ||\bm{r}_h - \operatorname{sg}[\bm{e}_{c_h}^h]||_2^2$.
$\hat{\bm{z}}$ is the reconstructed item embedding, and $\operatorname{sg}[\cdot]$ denotes the stop-gradient operation. 
$\beta$ is used to balance the optimization between the encoder and codebooks, typically set to 0.25.

\subsubsection{Semantically Relevant Tokenizers}
\label{sec:rel_tokenizers}
To obtain multiple item tokenizers that associate each item with multiple identifiers, a naive method involves training multiple RQ-VAE models, each initialized with different random parameters. 
However, the models learned in this operate independently, resulting in the token sequences tokenized by them being extraneous.
Thus, there is no related and homogeneous knowledge among the multiple groups of token sequence data constructed by these item tokenizers, which may even lead to serious semantic conflicts.

In contrast, we propose regarding the model checkpoints corresponding to adjacent epochs within a training process as multiple semantically relevant item tokenizers. 
These checkpoints are derived from iterative gradient descent from the identical initialization parameters, ensuring that disparities between codebooks in adjacent epochs remain minimal. 
The token sequences generated by these item tokenizers encapsulate related yet distinct semantic knowledge.
Formally, the learned multiple semantically relevant item tokenizers are written as:
\begin{align}
   \mathcal{T} & = \{\mathrm{T}_1, \mathrm{T}_2, \dots, \mathrm{T}_n\} \\
                & = \{\mathrm{T}_{\mathbf{\phi}^{N-n+1}}, \mathrm{T}_{\mathbf{\phi}^{N-n+2}} \dots, \mathrm{T}_{\mathbf{\phi}^{N}}\},
\end{align}
where $\mathcal{T}$ denotes a set of item tokenizers, and $n$ is the number of tokenizers. $\mathbf{\phi}^i$ represents the RQ-VAE parameter corresponding to the $i$-th epoch. $N$ indicates the maximum number of epochs.

\subsubsection{Tokenize an Item Sequence to Multiple Token Sequences}
\label{sec:tokenize_seq}

With the learned semantically relevant item tokenizers, a historical item sequence $S$ and a target item $v_{t+1}$ can be tokenized into multiple token sequences via different tokenizers:
\begin{align}
    X_1, X_2, \dots, X_n &= \mathrm{T}_1(S), \mathrm{T}_2(S),\dots, \mathrm{T}_n(S), \\
    Y_1, Y_2, \dots, Y_n &= \mathrm{T}_1(v_{t+1}), \mathrm{T}_2(v_{t+1}),\dots, \mathrm{T}_n(v_{t+1}),
\end{align}
where $X_i$ and $Y_i$ represent the token sequence and target item identifier tokenized by $\mathrm{T}_i$.
Notably, we do not directly utilize all augmented token sequences for model pre-training.
The reason is that when $n$ is large, the resulting data volume becomes unmanageable, and it is infeasible to adaptively adjust the proportions of different data groups.
Instead, we sample just one token sequence at a time for model optimization, which is approximately equivalent to using all the data through multiple sampling.
In the following section, we detail the method for adjusting the sampling probabilities of different data groups corresponding to the item tokenizers.

\subsection{Curriculum Recommender Pre-training}
\label{sec:data_curr}
Based on multi-identifier item tokenization, we obtain a data mixture involving multiple groups of token sequences, from which we select instances for generative recommender pre-training.
This presents a key challenge similar to that in LLM pre-training, namely how to adaptively adjust the proportions of different data groups during pre-training~\cite{llm_survey}. 
Inspired by the data curriculum proposed and widely employed in LLM pre-training~\cite{skill_it,DBLP:journals/corr/abs-2310-02263}, we devise a curriculum pre-training scheme based on data influence estimation in MTGRec.
Particularly, we estimate data influences corresponding to multiple item tokenizers via first-order gradient approximation (Section~\ref{sec:less}). 
Then, we dynamically adjust the sampling probabilities of different data groups in accordance with their estimated data influences for recommender pre-training (Section~\ref{sec:pre-train}).

\subsubsection{Estimating Data Influence}
\label{sec:less}
In order to more effectively utilize data from multiple item tokenizers, our idea is to increase the proportion of useful data while decreasing the proportion of low-quality data. 
To measure whether the data is ``\emph{useful}'' in a rational manner, we define the contribution of training data to the validation loss as data influence~\cite{sgd_estim,less,DBLP:conf/acl/HanSMTCW23}, and estimate it based on gradient information.
Formally, using the first-order Taylor expansion, the validation loss can be expressed as follows:
\begin{align}
    \mathcal{L}(\mathcal{D}_{val};\mathbf{\theta}^{t+1}) = \mathcal{L}(\mathcal{D}_{val};\mathbf{\theta}^{t}) + \nabla \mathcal{L}(\mathcal{D}_{val};\mathbf{\theta}^{t}) \cdot (\mathbf{\theta}^{t+1} - \mathbf{\theta}^{t}),
    \label{eq:taylor_expan}
\end{align}
where $\mathcal{D}_{val}$ denotes the held-out data for validation, and $\mathbf{\theta}^t)$ is the recommender parameter at time step $t$.
The first term of the equation represents the validation loss at time step $t$, while the second term is the first-order derivative within the Taylor expansion.
Then the update of validation loss is:
\begin{align}
    \mathcal{L}(\mathcal{D}_{val};\mathbf{\theta}^{t+1}) - \mathcal{L}(\mathcal{D}_{val};\mathbf{\theta}^{t}) = \nabla \mathcal{L}(\mathcal{D}_{val};\mathbf{\theta}^{t}) \cdot (\mathbf{\theta}^{t+1} - \mathbf{\theta}^{t}).
\end{align}

\paratitle{Calculate Gradient of Validation Data.}
Specific to the sequential recommendation scenario discussed in this paper, the validation data is acquired with the leave-one-out strategy.
After item tokenization using different tokenizers, multiple groups of token sequence data are mixed into $\mathcal{D}_{val}$.
The terms $\mathcal{L}(\mathcal{D}_{val};\mathbf{\theta})$ and $\nabla\mathcal{L}(\mathcal{D}_{val};\mathbf{\theta})$ denote the mean loss and cumulative gradient across all validation data, respectively, which can be formalized as:
\begin{align}
    \mathcal{L}(\mathcal{D}_{val};\mathbf{\theta}) &= \frac{1}{|\mathcal{D}_{val}|} \underset{{ X,Y \in \mathcal{D}_{val}}}{\sum} \mathcal{L}(X,Y;\mathbf{\theta}), \\
    \nabla\mathcal{L}(\mathcal{D}_{val};\mathbf{\theta}) & = \frac{1}{|\mathcal{D}_{val}|} \underset{{X,Y \in \mathcal{D}_{val}}}{\sum} \nabla \mathcal{L}(X,Y;\mathbf{\theta}), 
    \label{eq:grad_accu}
\end{align}
where $X, Y$ denotes a pair of token sequences corresponding to the historical interacted items and the target item. 
$\mathcal{L}(\cdot,\cdot ;\mathbf{\theta})$ is the negative log-likelihood loss in Eqn.~\eqref{eq:lmloss}.

\paratitle{Calculate Adam Gradients of Training Data.}
Since the generative recommender is usually trained using the Adam optimizer~\cite{adam}, the parameter update $\mathbf{\theta}^{t+1} - \mathbf{\theta}^{t}$ in Eqn.~\eqref{eq:taylor_expan} can be calculate as follows:
\begin{gather}
    \mathbf{\theta}^{t+1}-\mathbf{\theta}^{t} = -\eta_{t} \Gamma(\mathcal{D}^{i}_{train}; \mathbf{\theta}^{t}), \\
    \Gamma(\mathcal{D}^{i}_{train}; \mathbf{\theta}^{t}) = \frac{\bm{m}^{t+1}}{\sqrt{\bm{v}^{t+1}+\epsilon}}, \\
    \bm{m}^{t+1}  =(\beta_{1} \bm{m}^{t}+(1-\beta_{1}) \nabla \mathcal{L}(\mathcal{D}^{i}_{train}; \mathbf{\theta}^{t})) /(1-\beta_{1}^{t}), \\
    \bm{v}^{t+1} =(\beta_{2} \bm{v}^{t}+(1-\beta_{2}) \nabla \mathcal{L}(\mathcal{D}^{i}_{train}; \mathbf{\theta}^{t})^{2}) /(1-\beta_{2}^{t}),
\end{gather}
where $\mathcal{D}^{i}_{train}$ denotes the training token sequence data tokenized by the item tokenizer $\mathrm{T}_i$.
$\beta_1$ and $\beta_2$ are the hyperparameters of the first-order and second-order momentum in Adam, which are usually set to 0.9 and 0.999 respectively.
$\eta_{t}$ is the learning rate at time step $t$.
In our context, we do not estimate the influence of each individual data instance as in previous studies~\cite{sgd_estim,less,DBLP:conf/acl/HanSMTCW23}. 
Instead, we consider a group of data from each item tokenizer as an entirety for analysis. 
The gradient of each group of data (\ie $\nabla \mathcal{L}(\mathcal{D}^{i}_{train}; \mathbf{\theta})$) is calculated through gradient accumulation similar to Eqn.~\eqref{eq:grad_accu}, which is equivalent to treating $\mathcal{D}^{i}_{train}$ as a batch of data.

\paratitle{Calculate Influence.}
Based on the above analysis, we define the data influence of each item tokenizer at time step $t$ as:
\begin{align}
    \mathrm{I}(\mathrm{T}_i; \mathbf{\theta}^{t}) = \eta_{t} \nabla \mathcal{L}(\mathcal{D}_{val}; \mathbf{\theta}^{t}) \cdot \Gamma(\mathcal{D}^{i}_{train}, \mathbf{\theta}^{t}),
\end{align}
where $\mathrm{I}(\mathrm{T}_i; \mathbf{\theta}^{t})$ denotes the influence of the data group associated with the tokenizer $\mathrm{T}_i$.
Finally, given that the training process spans multiple time steps, we calculate the cumulative influence based on multiple model checkpoints as $\tilde{\mathrm{I}}(\mathrm{T}_i)=\sum_{k = 1}^K\mathrm{I}(\mathrm{T}_i;\mathbf{\theta}_k)$, where $\mathbf{\theta}_k$ indicates the $k$-th checkpoint at time step $t_k$ and $K$ denotes the total number of checkpoints.


\subsubsection{Curriculum Pre-training}
\label{sec:pre-train}
After elaborating on how to estimate the influence of the data from each item tokenizer, we now formulate a data curriculum scheme for model pre-training by dynamically adjusting the sampling probabilities of different data groups.
Specifically, we partition the training process into multiple stages, with each stage containing a specific number of epochs. 
At the end of each stage, we update data sampling probabilities based on the latest data influence of each item tokenizer, as determined by the current model checkpoint.
Formally, given the current model checkpoint $\mathbf{\theta_k}$ and the cumulative data influence $\tilde{\mathrm{I}}_{k-1}(\mathrm{T}_i)$ of the previous stage, the sampling probability is updated as follows:
\begin{align}
    \tilde{\mathrm{I}}_{k}(\mathrm{T}_i) &=\tilde{\mathrm{I}}_{k-1}(\mathrm{T}_i) + \mathrm{I}(\mathrm{T}_i; \mathbf{\theta}_k), \\
    p^{k}_i  &=  \frac{e^{  \tilde{\mathrm{I}}_{k}(\mathrm{T}_i) / \tau} }{\sum_{j=1}^n e^{ \tilde{\mathrm{I}}_{k}(\mathrm{T}_j) / \tau} }, 
    \label{eq:samp_p}
\end{align}
where $\tau$ denotes the temperature coefficient used to control the smoothness of the distribution, and $p^{k}_i$ is the sampling probability of the item tokenizer $\mathrm{T}_i$ for the subsequent stage. 
Initially, each data group is sampled with equal probability.
Then, the data sampling strategy for stage $k+1$ is defined as follows:
\begin{align}
   & \mathrm{T} \sim \mathcal{T} = \{\mathrm{T}_1, \mathrm{T}_2, \dots, \mathrm{T}_n\}, \\
   & P(\mathrm{T} = \mathrm{T}_i) = p^{k}_i, \\
   & X =\mathrm{T}(S), \quad Y = \mathrm{T}(v_{t+1}).
\end{align}
Finally, the sampled token sequence data $X$ and $Y$ are subsequently fed into the generative recommender for model optimization with the negative log-likelihood loss:
\begin{align}
    \mathcal{L}(X,Y) = - \sum_{h=1}^{H}\operatorname{log} P(c^{t+1}_h|X,c^{t+1}_1,\dots,c^{t+1}_{h-1}).
    \label{eq:lmloss}
\end{align}

\subsection{Fine-tuning and Inference}

\paratitle{Fine-tuning for Item Identification.}
In practical applications, the token sequence generated by the recommender should be able to identify the corresponding item. 
That is, the items and their identifiers should fulfill a one-to-one mapping within the recommender system. 
However, during our proposed curriculum pre-training with multiple item tokenizers, is unable to identify items since there might be multiple identifiers corresponding to the same item (\ie $\mathrm{T}_1(v), \dots, \mathrm{T}_n(v) \mapsto v$).
Therefore, we further fine-tune the pre-trained generative recommender based on each item tokenizer respectively and select the model with the optimal validation performance for actual deployment and testing.

\paratitle{Inference.}
Our objective during the inference phase is to generate the top $K$ items from the entire item set for recommendation. 
To achieve this, we adopt beam search to decode $K$ token sequences and map them to the corresponding items. 
Unlike some prior works~\cite{letter,howtoindex}, we do not introduce a prefix tree to constrain the search process, as it would hinder parallel decoding and reduce efficiency. 
As for invalid identifiers, which occur only rarely~\cite{tiger}, are simply ignored.

\section{Experiments}
\label{sec:experiments}

In this section, we conduct empirical experiments and in-depth analyses on three public datasets to demonstrate the effectiveness of our proposed MTGRec.

\subsection{Experiment Setup}

\begin{table}[]
    \centering
    \caption{Statistics of the preprocessed datasets. Avg.\textit{len} denotes the average length of item sequences.}
    \huge
    \resizebox{\linewidth}{!}{
    \begin{tabular}{lrrrrr}
    \toprule
     Dataset    &\#Users   &\#Items   &\#Interactions &Sparsity &Avg.\textit{len}  \\
     \midrule
     Instrument &57,439  &24,587  &511,836  &99.964\% &8.91 \\
     Scientific &50,985  &25,848  &412,947  &99.969\% &8.10\\
     Game &94,762  &25,612  &814,586  &99.966\% &8.60\\
     \bottomrule
    \end{tabular}}
    \label{tab:data_statistics}
\end{table}

\subsubsection{Dataset}
We evaluated the proposed approach on three subsets of the latest Amazon 2023 review dataset~\cite{amazon2023}, \ie ``\emph{Musical Instruments}'', ``\emph{Industrial and Scientific}'' and ``\emph{Video Games}''. These datasets contain user review data spanning from May 1996 to September 2023. 
In line with the preprocessing steps outlined in previous studies~\cite{s3rec,tiger}, we filter out low-activity users and items with less than five interaction records. 
Subsequently, we group the historical item sequences by users and sort them in chronological order, with a maximum sequence length limit of 20 items.
The detailed statistics of preprocessed datasets are presented in Table~\ref{tab:data_statistics}.

\subsubsection{Baseline Models}
To facilitate a comprehensive comparison, we categorize the baseline models into the following two groups:

\noindent \textbf{(1) Traditional sequential recommendation models}:
\begin{itemize}
\item {\textbf{Caser}}~\cite{caser} leverages convolutional neural networks to capture spatial and positional patterns in user behavior sequences.
\item {\textbf{HGN}}~\cite{hgn} uses feature-level and instance-level gating mechanisms to model user preference.
\item {\textbf{GRU4Rec}}~\cite{gru4rec} employs GRUs to capture sequential patterns in user interactions.
\item {\textbf{BERT4Rec}}~\cite{bert4rec} utilizes a bi-directional self-attentive model with mask prediction objective for sequence modeling.
\item {\textbf{SASRec}}~\cite{sasrec} adopts a unidirectional self-attention network for user behavior modeling.
\item {\textbf{FMLP-Rec}}~\cite{fmlp-rec} proposes an all-MLP model with learnable filters to reduce noise and model user preference.
\item {\textbf{HSTU}}~\cite{hstu} incorporates user actions and timestamps into the next item prediction and proposes hierarchical sequential transducers with significant scalability. Note that it is still an ID-based method.
\item {\textbf{FDSA}}~\cite{fdsa} introduces a dual-stream self-attention framework that independently models item-level and feature-level sequence for recommendation. 
\item {\textbf{S$^3$-Rec}}~\cite{s3rec} enhances sequential recommendation models by leveraging feature-item correlations as self-supervised signals.
\end{itemize}
\noindent \textbf{(2) Generative recommendation models}:
\begin{itemize}
\item {\textbf{TIGER}}~\cite{tiger} employs RQ-VAE to quantize the item embedding into semantic IDs serving as the item identifier and adopts the generative retrieval paradigm for sequential recommendation.
\item {\textbf{LETTER}}~\cite{letter} extends TIGER by integrating collaborative and diversity regularization into RQ-VAE.
\item {\textbf{TIGER++}}~\cite{tiger} employs representation whitening and exponential moving average (EMA) techniques to enhance codebook learning and improve the quality of semantic IDs. For implementation details, please refer to Section~\ref{sec:implement}.
\end{itemize}

\begin{table*}[]
\centering
\caption{The overall performance comparisons between different baseline methods and MTGRec. The best and second-best results are highlighted in bold and underlined font, respectively.}
\label{tab:main_res}
\huge
\resizebox{\textwidth}{!}{
\renewcommand\arraystretch{1.1}
\begin{tabular}{lcccccccccccc}
\toprule 
\multicolumn{1}{l}{\multirow{2}{*}{Methods}} & \multicolumn{4}{c}{Instrument}          & \multicolumn{4}{c}{Scientific}          & \multicolumn{4}{c}{Game}                \\
\cmidrule(l){2-5} \cmidrule(l){6-9} \cmidrule(l){10-13} 
\multicolumn{1}{l}{}                         & Recall@5 & Recall@10 & NDCG@5 & NDCG@10 & Recall@5 & Recall@10 & NDCG@5 & NDCG@10 & Recall@5 & Recall@10 & NDCG@5 & NDCG@10 \\
\midrule
Caser                                        & 0.0241   & 0.0386    & 0.0151 & 0.0197  & 0.0159   & 0.0257    & 0.0101 & 0.0132  & 0.0330    & 0.0553    & 0.0209 & 0.0281  \\
HGN                                          & 0.0321   & 0.0517    & 0.0202 & 0.0265  & 0.0212   & 0.0351    & 0.0131 & 0.0176  & 0.0424   & 0.0687    & 0.0271 & 0.0356  \\
GRU4Rec                                      & 0.0324   & 0.0501     & 0.0209 & 0.0266  & 0.0202    & 0.0338    & 0.0129 & 0.0173  & 0.0499  & 0.0799     & 0.0320  & 0.0416  \\
BERT4Rec                                     & 0.0307   & 0.0485    & 0.0195 & 0.0252  & 0.0186   & 0.0296    & 0.0119 & 0.0155  & 0.0460    & 0.0735    & 0.0298 & 0.0386  \\
SASRec                                       & 0.0333   & 0.0523    & 0.0213 & 0.0274  & 0.0259   & 0.0412    & 0.0150  & 0.0199  & 0.0535   & 0.0847    & 0.0331 & 0.0438  \\
FMLP-Rec                                     & 0.0339   & 0.0536    & 0.0218 & 0.0282  & 0.0269   & 0.0422    & 0.0155 & 0.0204  & 0.0528   & 0.0857    & 0.0338 & 0.0444  \\
HSTU                                         & 0.0343   & 0.0577    & 0.0191 & 0.0271  & 0.0271   & 0.0429    & 0.0147 & 0.0198  & 0.0578   & 0.0903    & 0.0334 & 0.0442  \\
FDSA                                         & 0.0347   & 0.0545    & 0.0230  & 0.0293  & 0.0262   & 0.0421    & 0.0169 & 0.0213  & 0.0544   & 0.0852    & 0.0361 & 0.0448   \\
S$^3$-Rec                                       & 0.0317   & 0.0496    & 0.0199 & 0.0257  & 0.0263   & 0.0418    & 0.0171 & 0.0219  & 0.0485   & 0.0769    & 0.0315 & 0.0406  \\
\midrule
TIGER                                        & 0.0370    & 0.0564    & 0.0244 & 0.0306  & 0.0264   & 0.0422    & 0.0175 & 0.0226  & 0.0559   & 0.0868    & 0.0366 & 0.0467  \\
LETTER                                       & 0.0372   & 0.0580     & 0.0246 & 0.0313  & 0.0279   & 0.0435    & 0.0182 & 0.0232  & 0.0563   & 0.0877    & 0.0372 & 0.0473  \\
TIGER++                                      & \underline{0.0380}    & \underline{0.0588}    & \underline{0.0249} & \underline{0.0316}  & \underline{0.0289}   & \underline{0.0450}     & \underline{0.0190}  & \underline{0.0241}  & \underline{0.0580}    & \underline{0.0914}    & \underline{0.0377} & \underline{0.0485}  \\
\midrule
MTGRec                                       & \textbf{0.0413}   & \textbf{0.0635}    & \textbf{0.0275} & \textbf{0.0346}  & \textbf{0.0322}   & \textbf{0.0506}    & \textbf{0.0212} & \textbf{0.0271}  & \textbf{0.0621}         &  \textbf{0.0956}         & \textbf{0.0410}       & \textbf{0.0517}        \\
Imporve   & +8.68\%    & +7.99\%     & +10.44\% & +9.49\%   & +11.42\%   & +12.44\%    & +11.58\% & +12.45\%  &   +7.07\%       &   +4.60\%        &  +8.75\%      &  +6.60\%  \\
\bottomrule
\end{tabular}
}
\end{table*}

\subsubsection{Evaluation Settings}

We adopt top-$K$ Recall and Normalized Discounted Cumulative Gain (NDCG), with $K$ set to 5 and 10, to evaluate the model performance in sequential recommendation.  
Following prior studies~\cite{sasrec,s3rec,tiger}, we apply the \emph{leave-one-out} strategy to split training, validation, and test sets. 
For each user interaction sequence, the final item he/she interacted is used as the test data, the second most recent item is used as the validation data, and all other items are used for training.
For the sake of rigorous comparison, we conduct full ranking evaluation over the entire item set rather than sample-based evaluation. Additionally, the beam size of autoregressive decoding is set to 50 for all generative recommendation models.

\subsubsection{Implementation Details}
\label{sec:implement}
In this part, we introduce the implementation details of the item tokenizer and generative recommender in MTGRec respectively.

\paratitle{Item Tokenizer.} In accordance with TIGER~\cite{tiger}, we leverage Sentence-T5~\cite{sentence-t5} to encode the textual information associated with each item as its semantic embedding. 
Subsequently, we learn an RQ-VAE model with 3 codebooks of size 256 and an extra codebook for collision handling.
In addition, the following three techniques are introduced to enhance codebook learning: 
(i) PCA with representation whitening~\cite{bert-whitening} is applied to enhance the quality of item semantic embeddings. 
(ii) As previous studies~\cite{lc-rec,mbgen}, a deeper MLP with hidden layer sizes of [2048, 1024, 512, 256] is utilized as the encoder and decoder in RQ-VAE. The codebook dimension is set to 128. 
(iii) Apply exponential moving averages (EMA) instead of gradient descent for codebook learning, which is more stable and effective~\cite{vq-vae}.
We denote the method after applying these techniques as TIGER++ and use the same techniques to learn RQ-VAE in our MTGRec.
The model is optimized by Adagrad optimizer over 10K epochs, employing a learning rate of 0.001 and a batch size of 2048.
We select the RQ-VAE checkpoints of the final $n$ epochs as the semantically relevant item tokenizers in our proposed approach, and $n$ is tuned between 5 and 30 with an interval of 5.

\paratitle{Generative Recommender.} We adopt T5~\cite{t5} as the backbone of the recommender, which has a model dimension of 128, an inner dimension of 512, 4 attention heads with a dimension of 64, and employs the ReLU activation function.
We tune the number of model layers $L$ within \{1, 2, 3, 4, 5, 6, 7, 8\}, and both the number of encoder layers and decoder layers are set to $L$.
We set the batch size on each GPU to 256 and use 4 GPUs to pre-train the model for 200 epochs on all datasets.
As for the number of epochs in each stage for curriculum pre-training, we first train 60 epochs for gradient feature warmup and then perform sampling probability update every 20 epochs
Furthermore, the temperature coefficient $\tau$ tuned in \{0.1, 0.3, 1.0, 3.0, 5.0, 10.0\}.
The AdamW optimizer is used for both pre-training and fine-tuning, with the learning rates set to 0.005 and 0.0002 respectively.
In addition, a cosine scheduler is utilized to adjust the learning rate. 

We implement all traditional sequential recommendation models based on RecBole~\cite{recbole}, which is a user-friendly open-source library for recommender systems.
For a fair comparison, we set the embedding dimension of all models to 128 and obtain the best performance through hyperparameter grid search.
For all generative baseline models, we use the same model architecture as our MTGRec and tune $L$ between 1 to 8.

\begin{table*}[]
\centering
\caption{Ablation study of our method in the Instrument and Scientific datasets.}
\label{tab:ablation}
\resizebox{0.9\textwidth}{!}{
\begin{tabular}{lcccccccc}
\toprule 
\multicolumn{1}{l}{\multirow{2}{*}{Methods}} & \multicolumn{4}{c}{Instrument}          & \multicolumn{4}{c}{Scientific}                \\
\cmidrule(l){2-5} \cmidrule(l){6-9}
\multicolumn{1}{c}{}                         & Recall@5 & Recall@10 & NDCG@5 & NDCG@10 & Recall@5 & Recall@10 & NDCG@5 & NDCG@10 \\
\midrule
(0) MTGRec                                   & \textbf{0.0413}   & \textbf{0.0635}    & \textbf{0.0275} & \textbf{0.0346}  & \textbf{0.0322}   & \textbf{0.0506}    & \textbf{0.0212} & \textbf{0.0271}  \\
(1) \  \emph{w/o} Data curriculum                    & 0.0406   & 0.0618    & 0.0268 & 0.0338  & 0.0312   & 0.0487    & 0.0205 & 0.0263  \\
(2) \  \emph{w/o} Relevant tokenizers                & 0.0350   & 0.0548    & 0.0226 & 0.0290   & 0.0249   & 0.0404    & 0.0158 & 0.0208  \\
(3) \  \emph{w/o} Pre-training                       & 0.0380    & 0.0571    & 0.0247 & 0.0309  & 0.0285   & 0.0443    & 0.0181 & 0.0236  \\
\bottomrule
\end{tabular}
}
\end{table*}

\begin{figure*}[]
\centering
\includegraphics[width=0.99\linewidth]{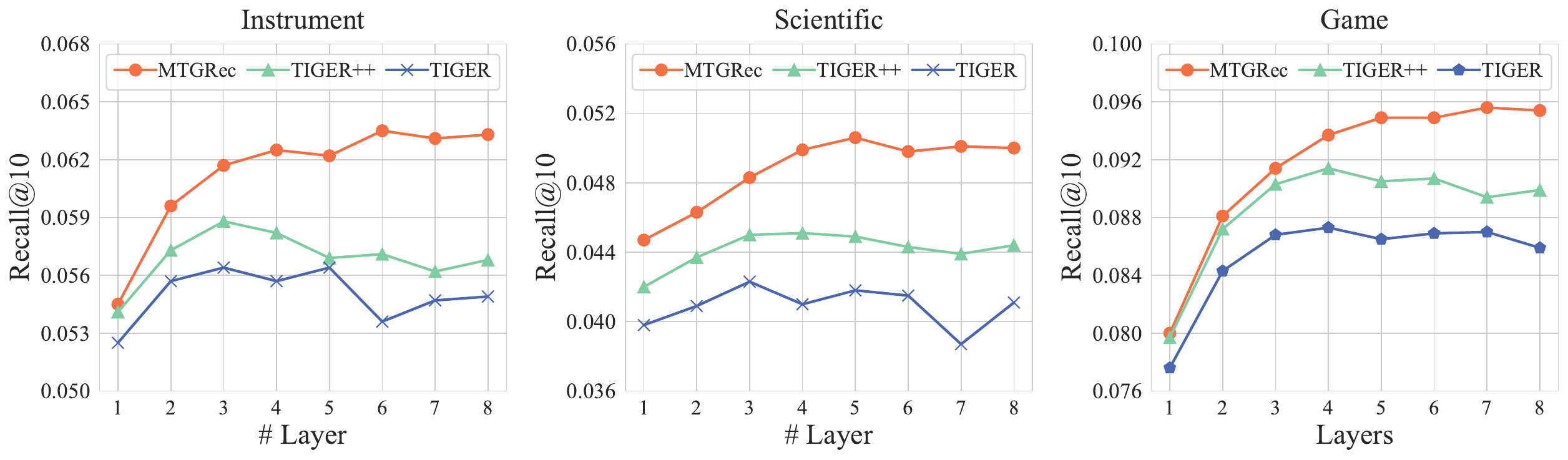}
\caption{Performance Comparison \wrt Model Scale. 
The x-axis coordinates are the number of encoder and decoder layers in the generative recommender. All reported results for MTGRec are the best results obtained using various numbers of tokenizers.}
\label{fig:model_scale}
\end{figure*}

\subsection{Overall Performance}
We compare MTGRec with both traditional and generative baseline models on three public recommendation benchmarks. The overall results are presented in Table~\ref{tab:main_res}. From these results, we can find:

For traditional sequential recommendation models, FMLP-Rec and HSTU attain better results than SASRec by introducing more advanced model architectures. 
S$^3$-Rec integrates auxiliary features for self-supervised pre-training, achieving outstanding results on the Game dataset.
Moreover, FDSA demonstrates superior performance compared to other models (\ie Caser, HGN, GRU4Rec, BERT4Re, SASRec, FMLP-Rec, HSTU) that only involve item ID and collaborative information on three datasets.
This observation implies that incorporating item textual features as supplementary information can significantly boost recommendation efficacy.

Regarding generative recommendation models, they generally outperform the traditional sequential recommendation models, benefiting from the item identifiers implying semantics and the generative paradigm.
Among them, LETTER and TIGER++ show better performance than TIGER, which is attributed to their improvements on the item tokenizer.
LETTER introduces collaborative and diversity regularization to integrate collaborative signals and alleviate code assignment bias for RQ-VAE.
TIGER++ applies representation whitening and EMA techniques to improve the item tokenizer in terms of item embedding quality and model optimization.

Finally, our proposed MTGRec consistently maintains the optimal performance in all cases, achieving substantial improvements over both traditional and generative baseline models.
Different from prior generative recommendation models, we introduce multiple semantically relevant item tokenizers for token sequence augmentation and design a model pre-training approach with data curriculum.
By pre-training the generative recommender on larger and more diverse sequence data derived from multiple item tokenizers, we significantly improve the model's scalability and effectiveness.

\subsection{Ablation Study}

To investigate the contributions of the various techniques included in our MTGRec, we conduct ablation studies on the Instrument and Scientific datasets and present the results in Table~\ref{tab:ablation}.
Specifically, we compare MTGRec with the following three variants:

(1) \underline{\emph{w/o Data curriculum}} without data curriculum based on influence estimation and samples data from different item tokenizers with equal probability.
We can see that this variant performs worse than MTGRec across all datasets, which indicates that introducing curriculum learning into generative recommender pre-training can effectively improve performance.

(2) \underline{\emph{w/o Relevant tokenizers}} learns multiple item tokenizers initialized with different random parameters for item tokenization, which are irrelevant and extraneous. 
The extraneous tokenizers cause serious semantic conflicts during pre-training, leading to the collapse of model learning and resulting in significant performance degradation.
This observation underscores the essence of selecting RQ-VAE checkpoints from adjacent epochs during a training process as semantically relevant item tokenizers.

(3) \underline{\emph{w/o} Pre-training} without pre-training on augmented sequence data derived from multiple semantically relevant item tokenizers, the generative recommender is learned based on a single item tokenizer (\ie TIGER++).
The results show that pre-training based on multiple item tokenizers serves as the critical element for the effectiveness of our framework.

\subsection{Further Analysis}

\subsubsection{Performance Comparison \wrt Model Scale}

In our framework, sequence augmentation via multiple item tokenizers provides us with more massive and diverse data.
The large-scale data motivates us to pursue enhanced performance via model scaling similar to that in LLMs.
Therefore, in this section, we endeavor to explore the influence of different model scales on recommendation performance. 
Specifically, we start from a single layer and progressively increase the number of encoder and decoder layers of the generative recommender up to 8 layers. 
For models of distinct scales, we experiment with different numbers of tokenizers for pre-training to attain optimal performance and present the results in Figure~\ref{fig:model_scale}. 
It is evident that MTGRec outperforms the baseline models (\ie TIGER, TIGER++) in all cases. 
Furthermore, the performance of baseline models is positively correlated with model scale only at shallow layers; as the model scale increases slightly (\eg with 4 or 5 layers), performance may degrade due to overfitting.
In contrast, the performance of MTGRec generally shows an upward tendency as the model scales. 
However, we acknowledge that this positive correlation is constrained, unlike in LLMs, where there remains room for improvement even as the model scale reaches 100B~\cite{llm_survey,scaling_law,scaling_instruct}.
This limitation may arise from the token sequence data being essentially constructed based on a finite set of observed user interactions.
As larger models require more data for effective optimization, it becomes challenging to achieve a trade-off between the quality and quantity of augmented data through multiple item tokenizers.
Specifically, the method fails to generate sufficient token sequences while maintaining the semantic relevance of data derived from RQ-VAE checkpoints separated by many training epochs. We leave addressing such issues to future work.

\begin{figure}[]
\centering
\includegraphics[width=0.99\linewidth]{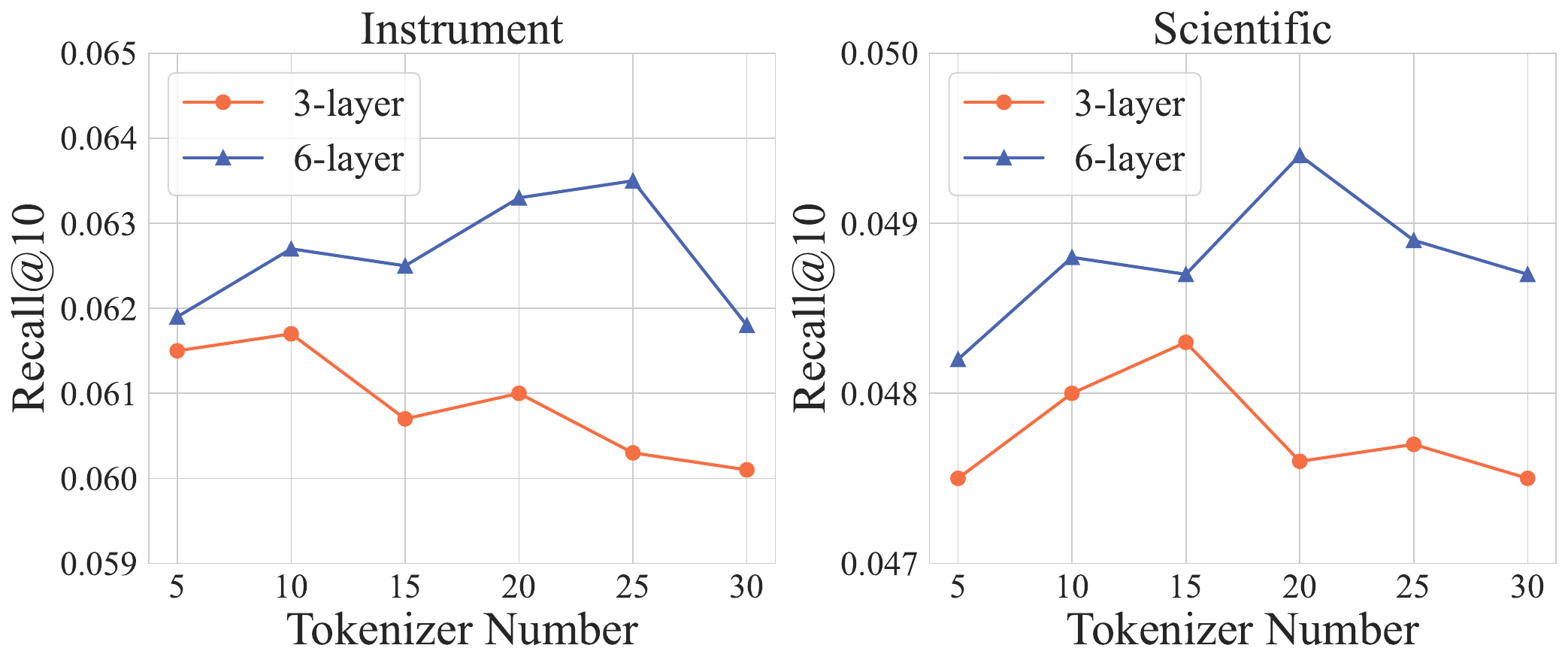}
\caption{Performance Comparison \wrt Tokenizer Number on the Instrument and Scientific datasets.}
\label{fig:token_num}
\end{figure}

\subsubsection{Performance Comparison \wrt Tokenizer Number}

In addition to model scale, we further investigate how the number of item tokenizers employed for model pre-training affects recommendation performance.
Specifically, we experiment with two scales of generative recommenders (\ie 3-layer and 6-layer models), pre-training them on datasets constructed with 5 to 30 item tokenizers.
As shown in Figure~\ref{fig:token_num}, pre-training with fewer tokenizers offers only marginal advancement, which can be attributed to the insufficient diversity and volume of sequence data for deep model optimization.
Furthermore, an excessively large number of item tokenizers will also lead to suboptimal performance. 
We hypothesize that when the intervals between item tokenizers span too many epochs, the semantic relevance between tokenizers weakens or even conflicts, thereby hindering effective model learning.  
Thus, it is crucial for MTGRec to select an appropriate number of item tokenizers to achieve a trade-off between data volume and semantic relevance. 
Besides,  we observe that the optimal number of tokenizers increases with model scale (\ie for 6-layer models), suggesting that larger models benefit from more extensive and diverse sequence data for effective training.

\begin{figure}[]
\centering
\includegraphics[width=0.99\linewidth]{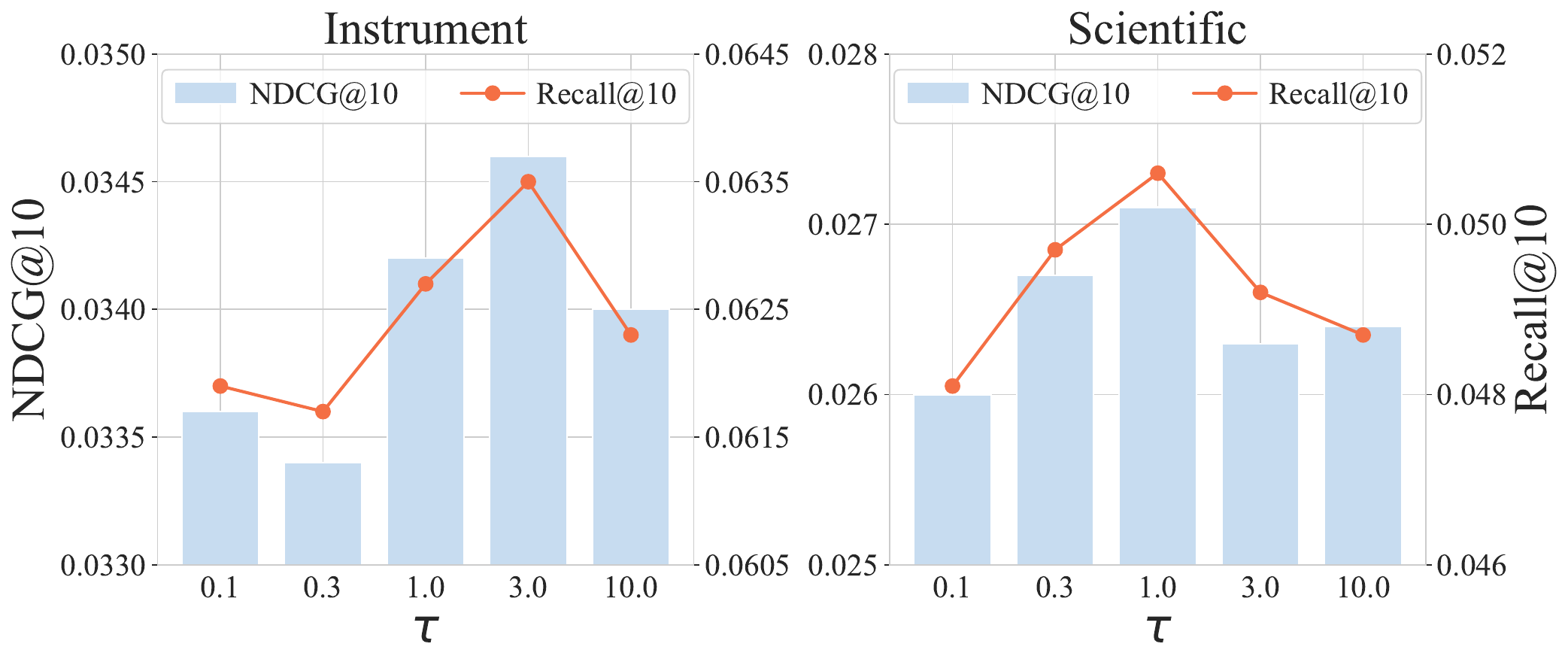}
\caption{Performance Comparison \wrt Temperature Coefficient on the Instrument and Scientific datasets. }
\label{fig:tau}
\end{figure}

\subsubsection{Performance Comparison \wrt Temperature Coefficient}

In the sampling probabilities of different data groups defined in Eqn.~\eqref{eq:samp_p}, the temperature coefficient $\tau$ is used to regulate the smoothness of the distribution. 
To evaluate the impact of $\tau$, we vary its value from 0.1 to 10 and report the results in Figure~\ref{fig:tau}. 
The results demonstrate that an appropriate $\tau$ can significantly improve the performance of MTGRec. 
Specifically, the optimal values of $\tau$ on the Instrument and Scientific datasets are 3 and 1, respectively.
A smaller $\tau$ leads the model to be more inclined towards high-probability item tokenizers while a larger $\tau$ results in the data curriculum degenerating into uniform sampling.
Both extremes adversely affect the effectiveness of curriculum pre-training.

\begin{figure}[]
\centering
\includegraphics[width=0.99\linewidth]{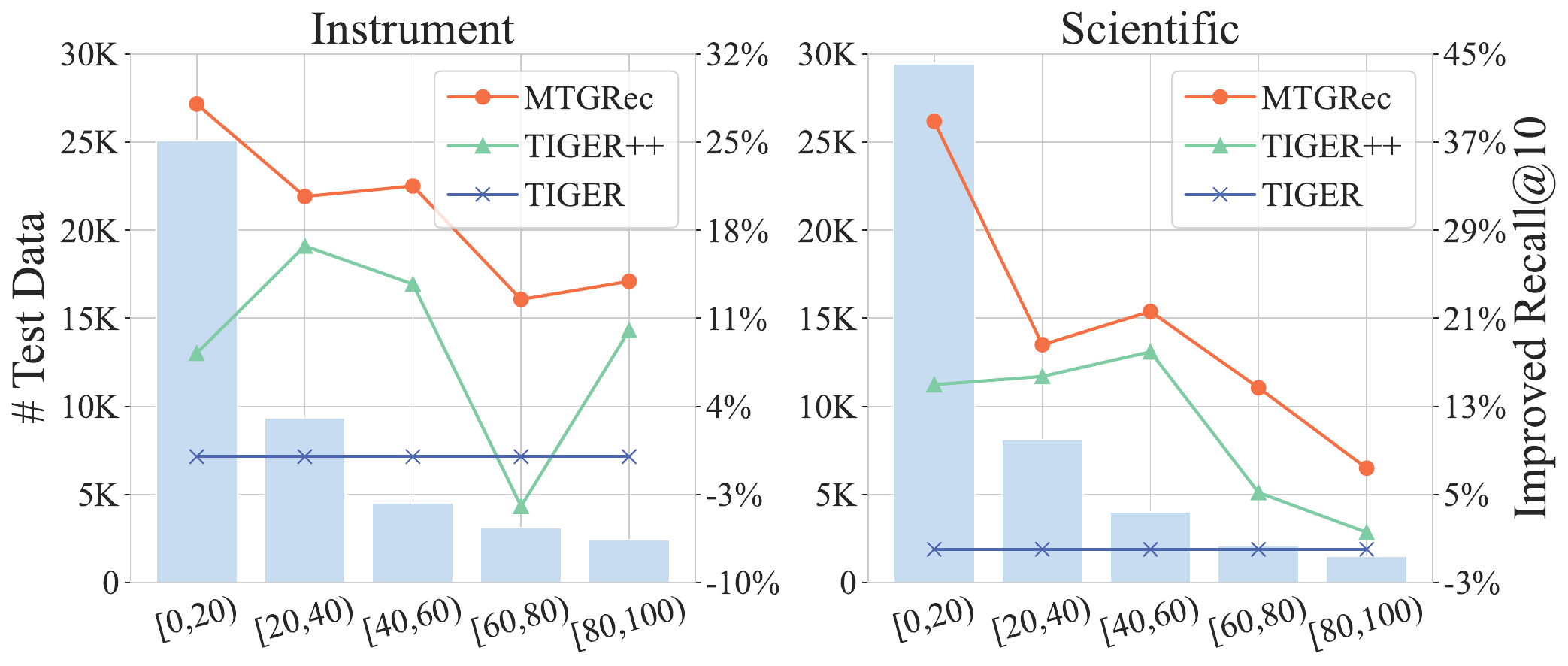}
\caption{Performance Comparison \wrt Long-tail Items on the Instrument and Scientific datasets.. The bar graph illustrates the number of interactions in the test data for each group, while the line chart displays the improvement ratios for Recall@10 in comparison to TIGER.}
\label{fig:long_tail}
\end{figure}

\subsubsection{Performance Comparison \wrt Long-tail Items}

One of the key motivations for developing a pre-training approach based on multiple item tokenizers is to enhance the generalization of the generative recommender and prevent it from disregarding long-tail items.
To verify the merit of our approach in recommendation involving long-tail items, we evaluate MTGRec on item groups with varying numbers of interactions.
Specifically, following prior works~\cite{unisrec}, we split the test data into different groups according to the popularity of target items and present the Recall@10 improvement over TIGER in Figure~\ref{fig:long_tail}.
We can see that MTGRec consistently outperforms the baseline model across all item groups. 
Especially when target items are unpopular, \eg group [0, 20), MTGRec shows superior performance and more significant improvement than TIGER and TIGER++. 
This phenomenon indicates that long-tail items can benefit from pre-training with multiple item tokenizers, as this approach provides increased exposure and incorporates more knowledge from shared tokens.

\begin{table}[]
\centering
\caption{Performance comparison on other generative recommendation methods. Our MTGRec significantly improves the performance of all models.}
\label{tab:other_method}
\huge
\resizebox{\linewidth}{!}{
\begin{tabular}{lcccc}
\toprule
\multicolumn{1}{c}{\multirow{2}{*}{Methods}} & \multicolumn{2}{c}{Instrument} & \multicolumn{2}{c}{Scientific} \\
\cmidrule(l){2-3} \cmidrule(l){4-5}
\multicolumn{1}{c}{}                         & Recall@10       & NDCG@10      & Recall@10       & NDCG@10      \\
\midrule
TIGER                                        & 0.0568          & 0.0307       & 0.0423          & 0.0225       \\
\ +MTGRec                                      & \textbf{0.0598}          & \textbf{0.0329}       & \textbf{0.0465}          & \textbf{0.0245}       \\
\midrule
LETTER                                       & 0.0580           & 0.0313       & 0.0435          & 0.0232       \\
\ +MTGRec                                      & \textbf{0.0614}          & \textbf{0.0335}       & \textbf{0.0481}          & \textbf{0.0255}       \\
\midrule
TIGER++                                      & 0.0588          & 0.0316       & 0.045           & 0.0241       \\
\ +MTGRec                                      & \textbf{0.0635}          & \textbf{0.0346}       & \textbf{0.0506}          & \textbf{0.0271}  \\
\bottomrule
\end{tabular}
}
\end{table}

\subsubsection{Applying MTGRec on Other Generative Recommendation Methods}

Furthermore, the proposed approach can be seamlessly integrated to other generative recommendation methods, such as the original TIGER and LETTER, with the only prerequisite being a trainable item tokenizer. 
To evaluate its general applicability, we applied MTGRec to additional generative recommendation methods on the Instrument and Scientific datasets.
As shown in Table~\ref{tab:other_method}, the results demonstrate that the proposed method can consistently improve the performance of the base models, further verifying its effectiveness. 
This confirms that selecting model checkpoints from adjacent epochs as semantically relevant item tokenizers the generation of sequence data with homogeneous knowledge across multiple methods.

\begin{table}[]
\centering
\caption{Item identifier difference \wrt different intervals.}
\label{tab:ids_diff}
\resizebox{\linewidth}{!}{
\begin{tabular}{lcccccc}
\toprule
\multirow{2}{*}{Intervals} & \multicolumn{2}{c}{Instrument} & \multicolumn{2}{c}{Scientific} & \multicolumn{2}{c}{Game} \\
\cmidrule(l){2-3} \cmidrule(l){4-5} \cmidrule(l){6-7}
                         & First       & Any     & First       & Any & First       & Any      \\
\midrule
1         & 0.39\%     & 13.58\% & 0.27\%    & 11.4\%   & 0.36\%    & 9.36\%   \\
5         & 0.44\%     & 21.26\% & 0.58\%    & 22.54\%  & 0.58\%    & 21.22\%  \\
10        & 0.51\%     & 29.75\% & 0.51\%    & 30.68\% & 0.57\%    & 30.43\%  \\
20        & 0.75\%     & 44.09\% & 0.71\%   & 47.33\% & 0.79\%   & 47.42\% \\
30        & 0.87\%     & 54.94\% & 0.85\%   & 58.29\% & 1.14\%   & 59.95\% \\
\bottomrule
\end{tabular}
}
\end{table}

\subsubsection{Multiple Identifier Difference Analysis}
In this section, we analyze the relevance and differences of item identifiers generated by item tokenizers at different training epochs. 
Specifically, given two sets of item identifiers generated by two item tokenizers, we calculate two metrics: (1) the proportion of items whose first token changed, and (2) the proportion of items with any token changes in the item identifier.
The results are shown in Table~\ref{tab:ids_diff}. 
We observe that for two adjacent tokenizers (\ie with an epoch interval of 1), the item identifiers on the three datasets exhibit minimal changes, indicating strong semantic consistency. 
As the epoch interval between tokenizers increases, a larger number of item identifiers change (\eg 59.95\% in the Game dataset), leading to a higher risk of semantic conflict. 
It is worth noting that even when the interval is large and many identifiers differ, the proportion of changes in the first token typically remains below 1\%, thereby preserving the core semantic information.

\begin{table}[]
\centering
\caption{Efficiency of different Methods.}
\label{tab:efficiency}
\resizebox{\linewidth}{!}{
\begin{tabular}{lcccccc}
\toprule
\multirow{2}{*}{Intervals} & \multicolumn{2}{c}{Instrument} & \multicolumn{2}{c}{Scientific} & \multicolumn{2}{c}{Game} \\
\cmidrule(l){2-3} \cmidrule(l){4-5} \cmidrule(l){6-7}
                         & Time       & Epoch     & Time       & Epoch & Time       & Epoch      \\
\midrule
TIGER   & 1.33 h & 186   & 1.04 h & 184   & 2.19 h  & 253   \\
TIGER++ & 1.22 h & 178   & 1.02 h & 187   & 2.23 h & 264   \\
MTGRec  & 1.41 h & 209   & 1.21 h & 217   & 2.11 h & 248 \\
\bottomrule
\end{tabular}
}
\end{table}

\subsubsection{Efficiency Analysis}
In this section, we further investigate the efficiency of the proposed method. 
As shown in Table~\ref{tab:efficiency}, we measure the training time and the number of epochs required for model convergence across different methods under the same settings and the same hardware environment. 
Our method, MTGRec, consists of a pre-training phase of 200 epochs, followed by a low-cost fine-tuning phase.
The results demonstrate that our multi-identifier pre-training strategy does not introduce excessive training time costs compared to baseline methods, while achieving significant performance gains.
Furthermore, the proposed curriculum learning scheme accelerates model convergence on the Game dataset.

\section{Related Work}
\label{sec:related}

In this section, we review related work from two aspects, namely, sequential recommendation and generative recommendation.

\subsection{Sequential Recommendation}

Sequential recommendation~\cite{sasrec,gru4rec} aims to capture user preferences based on the historical behavior sequence and predict the next item that the user is most likely to interact with.
Early studies~\cite{fpmc,fossil} employ the Markov Chain to model item sequences and learn the transformation relationship between items.
With the rapid advancements in deep learning, deep neural networks have emerged as a powerful tool for sequence modeling.
Therefore, recent works propose various sequential recommenders based on neural networks, including convolutional neural network (CNN)~\cite{caser}, recurrent neural network (RNN)~\cite{gru4rec, DBLP:conf/recsys/TanXL16}, graph neural network (GNN)~\cite{DBLP:conf/aaai/WuT0WXT19,DBLP:conf/ijcai/XuZLSXZFZ19}, and Transformer~\cite{sasrec,bert4rec}.
However, these approaches are primarily developed based on item IDs and collaborative filtering relations, while overlooking the wealth of information embedded in item content (\ie title, description, category).
More recently, there have been several attempts~\cite{fdsa,s3rec} that leverage additional information related to items to enhance ID sequence modeling.
Furthermore, pre-trained language models have been widely utilized to encode item textual features as semantic embeddings to improve performance and generalization~\cite{unisrec,vqrec,recformer}.

\subsection{Generative Recommendation}
Generative recommendation~\cite{tiger,ttds,siit,sc,mender}, as an emerging and promising paradigm,  has demonstrated superior performance on sequential recommendation tasks compared to traditional recommender systems.
In the generative paradigm, each item is indexed with an identifier represented by a list of tokens.
This process, known as item tokenization, plays a critical role in generative recommendation.
Existing item tokenization methods can be broadly categorized into three groups: heuristic approach, text-based approach, and codebook-based approach.
Heuristic approach mainly relies on manually defined rules or techniques, such as time order~\cite{howtoindex}, item clustering~\cite{seater, eager}, and matrix decomposition~\cite{gptrec, howtoindex}, to construct item identifiers. 
While these methods are easy to implement, they often fail to capture the implicit relationships between items, limiting their effectiveness.
Text-based approach directly utilizes item attributes, such as title, features, and description as identifiers~\cite{gpt4rec,DBLP:conf/recsys/Palma23,DBLP:conf/recsys/HarteZLKJF23,actionpiece}. 
These methods are usually designed to leverage the internal knowledge of pre-trained language models to improve recommendation performance. 
However, they suffer from problems such as inconsistent length, semantic ambiguity, and a lack of collaborative information.
In contrast, codebook-based approach~\cite{tiger,ding2024inductive,DBLP:journals/corr/abs-2403-18480,tokenrec} adopt learnable codebooks to quantize item embeddings, thereby constructing fixed-length, semantically rich item identifiers.
In addition, several recent studies have focused on enhancing codebook learning to better adapt it to recommendation systems. Notable examples include the introduction of collaborative and diversity regularization~\cite{letter}, as well as alignment between the item tokenizer and the generative recommender~\cite{etegrec}.

Reviewing the existing generative recommendation methods, most of them establish a one-to-one mapping between items and their identifiers, which results in challenges such as long-tail distribution, data sparsity, and insufficient diversity in token sequence data.
In contrast, in this paper, we introduce multiple semantically relevant item tokenizers to construct more massive and diverse data for generative recommender pre-training, aiming to enhance model scalability and performance.

\section{Conclusion}
\label{sec:conclusion}

In this paper, we introduced MTGRec, a framework that leverages multi-identifier item tokenization for generative recommender pre-training. 
Compared with previous methods that establish a \emph{one-to-one mapping} between items and their identifiers, MTGRec incorporates multiple item tokenizers to associate each item with several identifiers.  
Specifically, we first detailed the concept of multi-identifier item tokenization and then enhanced the generative recommender through curriculum pre-training.
For multi-identifier item tokenization, we proposed using RQ-VAE checkpoints corresponding to adjacent epochs as semantically relevant item tokenizers. 
These tokenizers enable the augmentation of item sequence data into multiple groups of token sequences, each with related but distinct semantic distributions.  
For curriculum recommender pre-training, we designed a data curriculum scheme based on data influence estimation to dynamically adjust the sampling probabilities of different data groups.
Finally, to ensure accurate item identification during recommendation, we fine-tuned the pre-trained model on each item tokenizer and selected the best model for deployment and testing.  
Extensive experiments and in-depth analyses on three public datasets demonstrated the superior performance of our proposed framework over both traditional and generative recommendation baselines.

For future work, we will adapt the multi-identifier item tokenization to more generalized recommendation scenarios such as transferable recommendation and multi-domain recommendation. Additionally, we will attempt to further scale the model parameters to the billion level and investigate the scaling effect when the model parameters are increased.


\bibliographystyle{ACM-Reference-Format}

\bibliography{ref}

\end{document}